\documentclass[runningheads]{llncs}
\usepackage{graphicx}
\usepackage{subcaption}
\usepackage{float}
\usepackage {amsmath}
\begin{document}
\title{ DeltaFinger: a 3-DoF Wearable Haptic Display Enabling High-Fidelity Force Vector Presentation at a User Finger \thanks{The reported study was funded by RFBR and CNRS, project number 21-58-15006.}}
\titlerunning{ DeltaFinger}
\author{Artem Lykov\orcidID{0000-0001-6119-2366} \and
Aleksey Fedoseev\orcidID{0000-0003-2506-9111} \and
Dzmitry Tsetserukou\orcidID{0000-0001-8055-5345}}
\authorrunning{A. Lykov et al.}
\institute{Skolkovo Institute of Science and Technology, Bol'shoy Bul'var, 30. 1, Moscow, Moscow Oblast, 121205, Russia\\
%\url{https://www.skoltech.ru/en/} \\
\email{\{artem.lykov, aleksey.fedoseev, d.tsetserukou\}@skoltech.ru}}
\maketitle % typeset the header of the contribution
\begin{abstract}
This paper presents a novel haptic device, named DeltaFinger, designed to deliver the force of interaction with virtual objects by guiding the user's finger by a wearable delta mechanism.

DeltaFinger delivers a 3D force vector to the fingertip of the index finger of the user, allowing complex rendering of various virtual reality (VR) environments. The developed device is able to render linear forces up to $1.8$ N in vertical projection and $0.9$ N in horizontal projection without restricting the motion freedom of the remaining fingers.

The experimental results showed a sufficient precision in perception of force vector with DeltaFinger (mean angular error in the perceived force vector of $0.6$ rad). The proposed device potentially can be applied to VR communications, medicine, and navigation for people with vision problems.

\end{abstract}

\begin{figure}[h!]\centering
\subfloat[]{\label{a}\includegraphics[width=.5\linewidth]{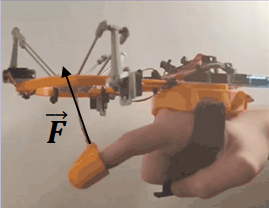}}\hfill
\subfloat[]{\label{b}\includegraphics[width=.5\linewidth]{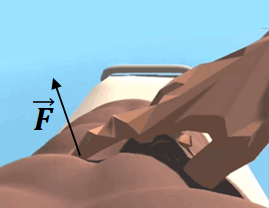}}\
\caption{(a) Kinesthetic feedback delivered by haptic display. (b) Force vector rendering in VR environment.}
\label{fig6}
\end{figure} 

\section{Introduction}
During the last decade, extensive research has been done in the field of wearable haptic interfaces for virtual reality (VR). Such interfaces allow their users a substantial benefit of the high mobility in VR environments and various modalities of interaction with virtual objects, thus, expanding the scope of applications with haptic feedback. For example, applications with wearable devices are beneficial in VR simulators for medical training, virtual CAD assembly, or remote control of robots through VR interfaces. The recent reviews of Pacchierotti et al. \cite{Pacchierotti_2017}, Cao et al. \cite{Cao_2018}, and See et al. \cite{See_2022} suggest that the majority of the developed devices are focused on providing kinesthetic feedback to the fingertips of users due to the large number of skin sensors located at this area.

However, methods to generate kinesthetic haptic stimuli on human fingertips remain to be further investigated. While a high number of papers proposed solutions for realistic and intuitive perceptual clues of the force amplitude and distribution, there is a lack of means to provide the directional cues, supporting user exploration of the VR environment. 

In this paper, we propose a novel wearable haptic device with kinesthetic haptic feedback delivered by a delta mechanism to render a force vector to be intuitive and recognizable by the user (Fig. \ref{fig6}). The developed interface delivers a linear force to the index finger of the user with a high range of force vector angles, allowing complex rendering of the VR environment.

%Haptic applications for virtual and augmented reality have been extensively explored over the last decade. Although many devices have been developed to transfer haptic sensations from VR, the task of transferring the force of interaction with VR objects remains unsolved. 

%Transferring interaction forces would extend the application of VR to the areas of industry, science, and communication, which require more information about the interaction. Examples of such applications include VR simulators for medical training, remote control of robotic manipulators using VR, and testing prototype models in VR. In all these cases, force transfer refers to the reflection of the force vector arising in VR in reality. This paper focuses on the development of a device to transfer a force vector to the index finger of the hand.

\section{Related Works}

Humans fingers are most often used for probing, grasping, sliding and otherwise manipulating both with real and virtual surfaces. Therefore, there is a number of methods designed for rendering particular haptic experiences during manipulation with virtual objects through wearable interfaces.

Several works explored force vector rendering through the indentation of the moving tactors into the skin. For example, Gabardi et al.\cite{Gabardi_2016} proposed the Haptic Thimble display with two rotations and one translation degrees of freedom (DoF) to render local orientation of the virtual surface. The similar approach was implemented by Benko et al. \cite{Benko_2016} with the NormalTouch handheld controller rendering up to 45 deg surface tilt by a 3-DoF Stewart Platform. Another approach was introduced by Tsetserukou et al. \cite{Tsetserukou_2014} with the LinkTouch interface where kinesthetic feedback is delivered by an inverted five-bar mechanism. This concept was later extended by Ivanov et al. \cite{Ivanov_2020} in the LinkRing device able to render force at two contact points of the human fingertip independently. The mentioned above displays allow to render with high resolution the point of applied force. However, the location of the actuators and the small area of the human fingertip provide additional challenges for a human perception of the force direction. Moriyama et al. \cite{Moriyama_2019} proposed to combine the force vector direction feedback from an inverted five-bar mechanism located on the forearm of the user with vibrotactile feedback of force amplitude delivered to the fingers. This approach improves surface orientation perception by providing feedback to the larger area of the human forearm. However, its naturalness should be further investigated.

Normal and shearing force vector rendering with a belt placed in contact with the user's fingertip skin was proposed by Minamizawa et al. \cite{Minamizawa_2007}. Pacchierotti et al. \cite{Pacchierotti_2016} suggested the hRing device aiming at realistic interaction without disturbance of hand tracking by locating belt on the proximal finger phalanx. The combination of shear force vector rendering with rotational platform and normal force vector rendering through electro-tactile display was introduced by Yem et al. \cite{Yem_2016} in FingAR interface. However, the mentioned above approaches are able to render only two horizontal directions of the haptic force aligned with the rotation direction of the actuators.

Exoskeletons mainly considered as heavy and cumbersome wearable haptic systems, with low naturalness and effectiveness of perceptual clues. However, a number of works are seeking a way to reduce the negative impact of exoskeleton designs. For example, Solazzi et al. \cite{Solazzi_2010} introduced finger exoskeleton for contact and orientation rendering. Agarwal et al. \cite{Agarwal_2015} proposed an exoskeleton delivering feedback at the index finger phalanx for post-stroke rehabilitation. Hernandez-Santos et al. \cite{Hernandez-Santos_2021} later suggested a finger exoskeleton able to provide haptic feedback to several phalanges, though only in 1-DoF. Li et al. \cite{Li_2022} explored the restraints of index finger motion for an optimized rehabilitation exoskeleton. Iqbal et al. \cite{Iqbal_2015} proposed a lightweight four-finger exoskeleton adjusting to variable hand sizes and other distinguishing features. More recently, Dragusanu et al. \cite{Dragusanu_2022} developed a lightweight modular exoskeleton that apply bidirectional forces. This design, however, is limited in force orientation rendering. Fang et al. \cite{Fang_2020} proposed a novel Wireality design with force vector rendering through strings pulling user's fingers. The actuators are located on the user's shoulder, which allows the device to apply higher forces at a cost of limiting force vector angles. Sim et al. \cite{Sim_2021} developed a low-latency exoskeleton glove allowing the adduction and abduction motions of each finger.

\section{System Overview}
\begin{figure}[h!]
\includegraphics[width=\textwidth]{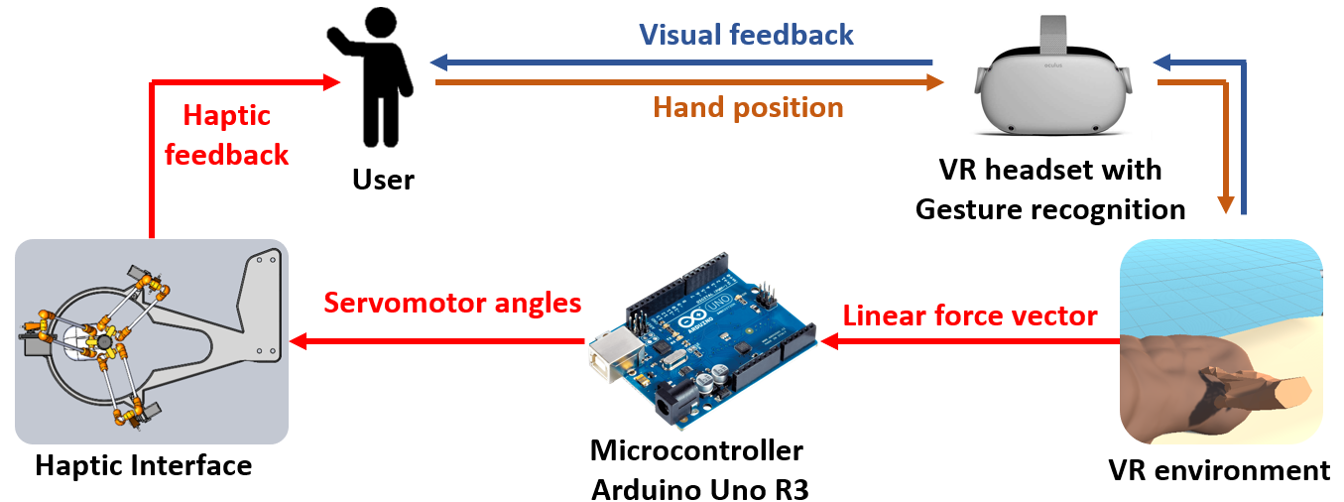}
\caption{System architecture. When user interacts with an object in the VR environment, the virtual framework calculates the position of their index finger with Oculus headset and renders the linear force vector at a point of a contact through the DeltaFinger wearable interface located above the user's index finger. } \label{fig1}
\end{figure}

The system architecture of the developed haptic interface is shown in Fig. \ref{fig1}.

The DeltaFinger hardware consists of the haptic interface based on delta mechanism driven by three TGY-TS531A analog nano servo motors, Arduino Uno R3 WiFi microcontroller, and Oculus Quest 2 VR headset. The system software consists of the VR framework developed on Unity Engine, that estimates force amplitude and direction, and middleware, that calculates motor angles based on the high level commands from Unity. The motor angles are calculated by inverse kinematic algorithm discussed in section \ref{sec_kin}. 

The operator is able to interact with virtual surfaces using the Oculus Quest 2 hand tracking interface. The virtual framework is developed to render the amplitude and direction of the linear force at the point of contact between users' hands and the surface of virtual object based on the approach proposed by Fedoseev et al. \cite{Fedoseev_2019}. Thus, when the user interacts with objects in VR,  a linear force vector is calculated at the point of their index finger collision with the object's surface. This vector is then delivered to the operator's hand via the DeltaFinger haptic interface.

%\subsection{Haptic Interface Design}

\subsection{DeltaFinger Kinematics}
\label{sec_kin}
%The mechanism to be designed should be small and lightweight, as it is intended to be worn on the arm and, if possible, not to be felt. 

The DeltaFinger interface design is based on a delta mechanism proposed by Trinitatova et al. \cite{Trinitatova2019} for delivering haptic feedback to human palm. The advantages of the parallel structure are in high load capacity with respect to the total weight of the device, low inertia, relatively high rigidity, and high speed. These advantages are caused by the multiple kinematic chains linking the end-effector and the frame of the device. A three-revolute (3-RRR) parallel structure is developed with three RRR serial chains that join in a fixed base. Each RRR chain is a serial chain composed by three rotational joints. The solution of the inverse kinematic problem for 3-RRR parallel structure was developed based on the paper of Ouafae et al. \cite{kinematics}.

\begin{figure}[H]\centering
\subfloat[]{\label{a}\includegraphics[width=.8\linewidth]{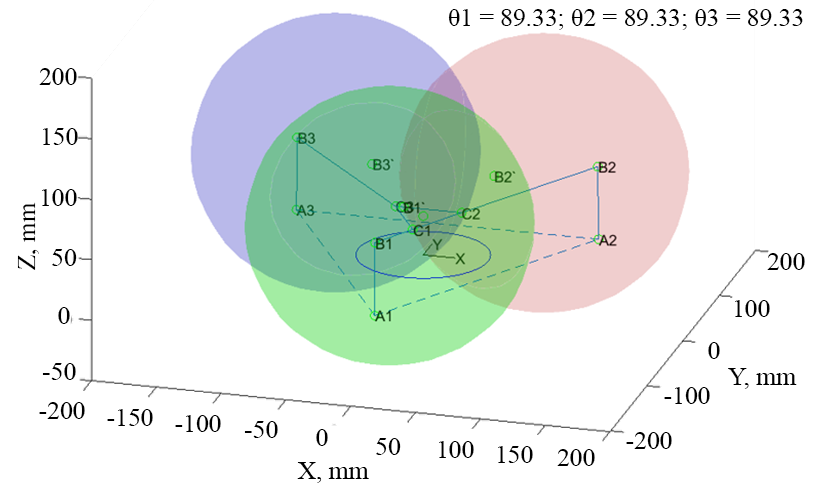}}\hfill
\subfloat[]{\label{b}\includegraphics[width=.8\linewidth]{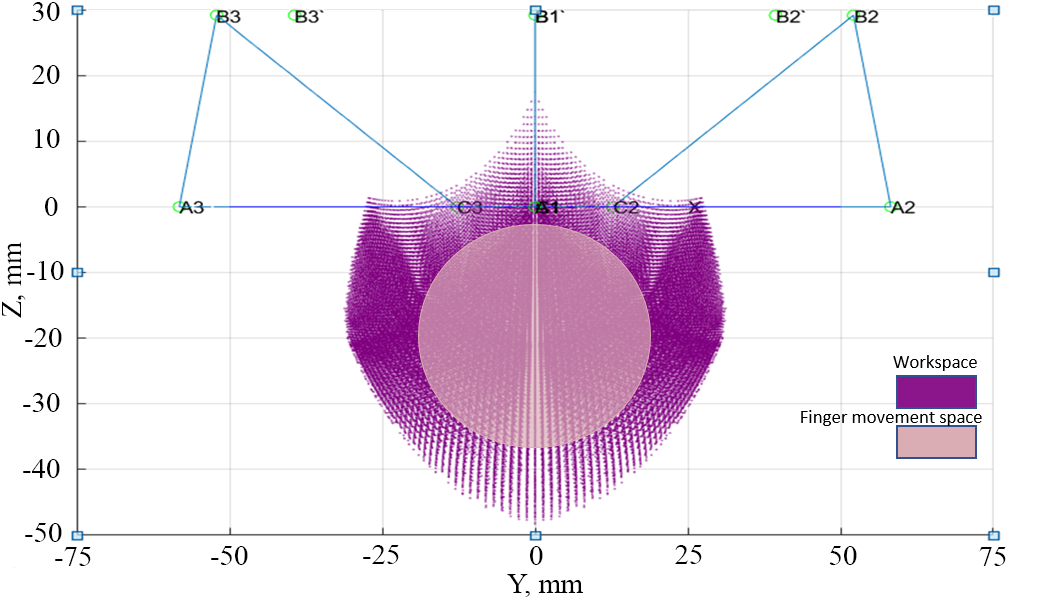}}\
\caption{DeltaFinger kinematics: (a) RRR chain plot. (b) Workspace of the end-effector.}
\label{fig2}
\end{figure}

The simulation in MATLAB Simulink was conducted to design the device with a working space that includes all points reachable by the motion of a human index finger, thus not restricting the mobility of the user's hand. The reachability of the three RRR chains and the resulting workspace of the DeltaFinger along with the motion space of the index finger is shown in Fig. \ref{fig2}. 

The circuit consists of three consecutive RRR kinematic chains that connect it to a fixed base with a radius of 80 mm. Each RRR chain is a sequential structure with three rotational joints connected by the links with the sizes of 35 mm, 60 mm, and 10 mm starting from the base.
The simulation results suggested that the DeltaFinger end-effector workspace is sufficient to cover an area with the radius of 25 mm, thus, allowing free motion of the user's finger in the YZ coordinate plane.

\begin{figure}[H]\centering
\includegraphics[width=0.8\textwidth]{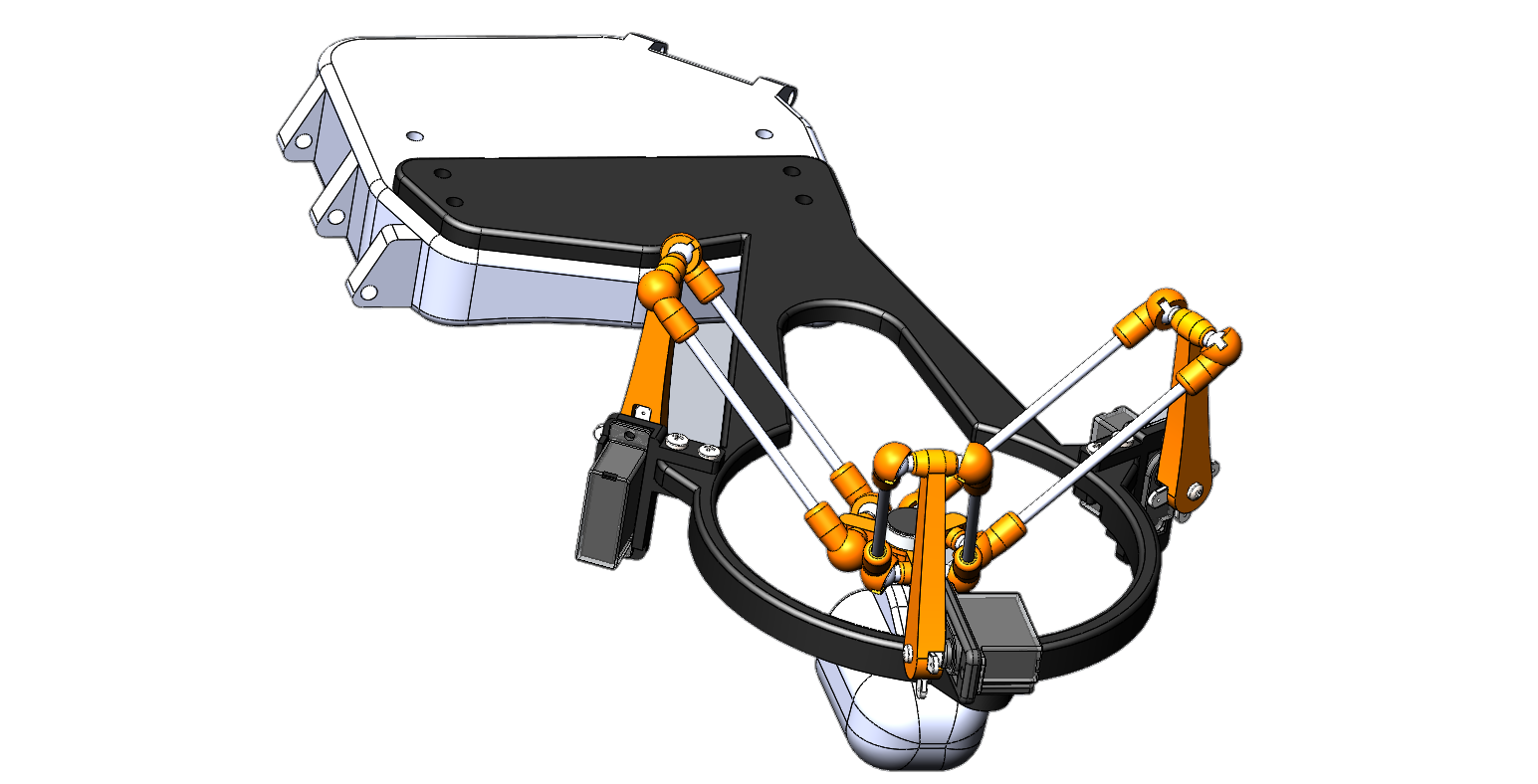}
\caption{DeltaFinger CAD design.} \label{fig4}
\end{figure}

The CAD model of the proposed device was designed in SolidWorks 2020 (Fig. \ref{fig4}). 

In addition, the base is fixed to the operator's hand by two flexible belts, and the output link is fixed to an index finger by a thimble with elastic inner surface.

\section{Haptic Rendering}
\begin{figure}[H]\centering
\includegraphics[width=0.7\textwidth]{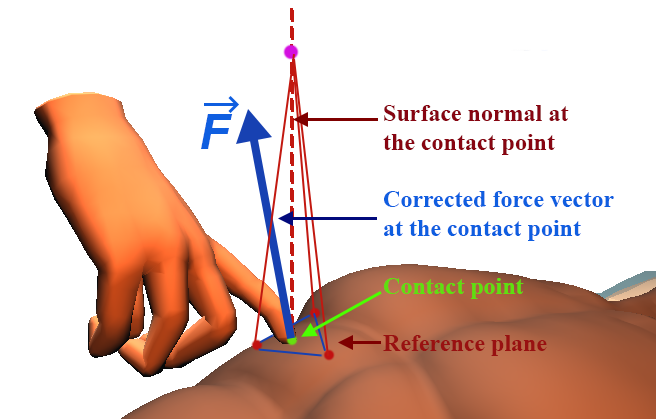}
\caption{Force vector rendering in virtual environment.} \label{fig5}
\end{figure}

The performance of the force vector rendering algorithm is shown in Fig. \ref{fig5}. 

To ensure the smooth transition of the force vector orientation and, thus, the stable performance of the inverse kinematics algorithm of the device, the direction of the surface normal was estimated not on the single point of collision but with the supporting plane defined by the three reference points at the "contact patch", i.e., 50x50 mm area around the collision point. To calculate the plane orientation, firstly, we calculate the normal vector to the surface at the contact point. Then, three rays are released from the point that was selected on this normal vector at 10 cm above the finger. The rays are faced towards the virtual surface at a 15 deg angle to the normal vector. The reference points are then obtained as the points of collision between the rays and the virtual surface. These points define the reference plane in 3D Cartesian space. Finally, the interaction force vector is defined as a normal vector to the reference plane. The amplitude of the force is proportional to the distance between the finger and the reference plane. The force is then transmitted to the Arduino microcontroller board and delivered to a user's finger by the developed haptic interface. 

% \begin{figure}[h!]
% \centering
% \includegraphics[width=0.5\textwidth]{fig6.png}
% \caption{Force transfer example} \label{fig6}
% \end{figure}

\section{Experimental Evaluation}
\subsection{Force Vector Rendering Experiment}
\begin{figure}[h!]
\includegraphics[width=\textwidth]{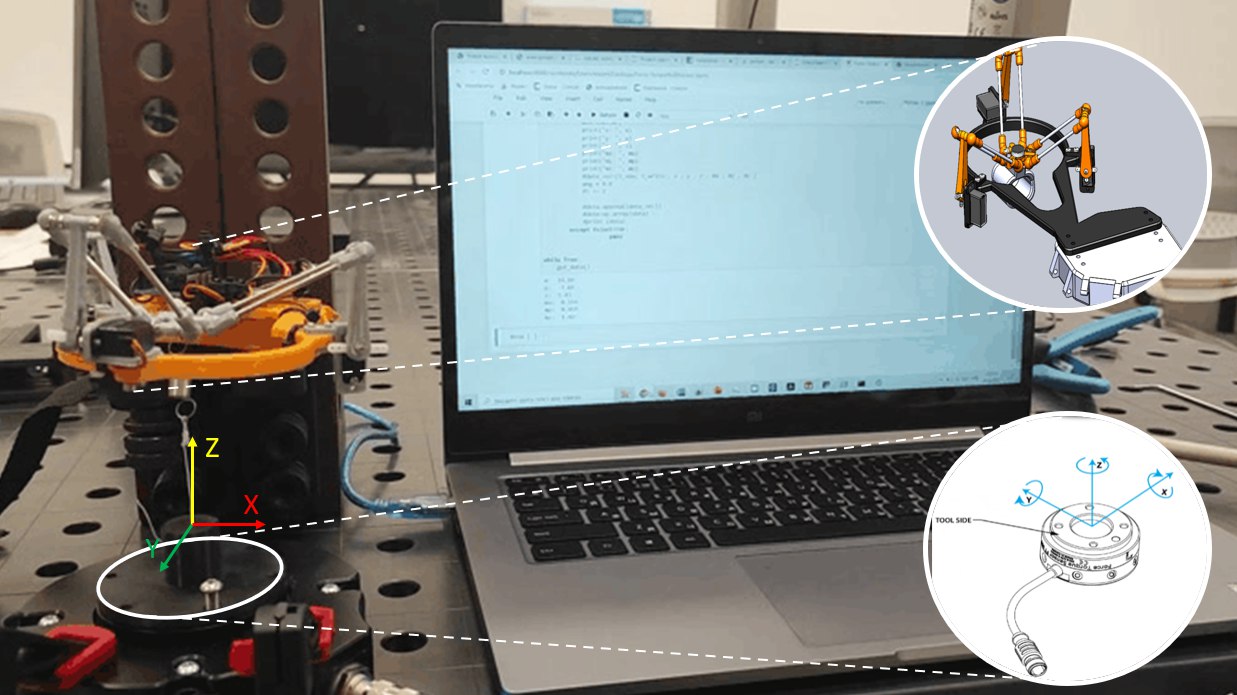}
\caption{Experimental setup for evaluation of the FigngerGuider applied forces with Robotiq 6-DoF force sensor.} \label{fig7}
\end{figure}

The experiment was carried out to estimate whether the maximal forces at the end-effector were adequate to render a virtual surface. For this, an experimental setup was assembled to evaluate the applied force by the Robotiq 6-DoF force/torque sensor \cite{Sensor}. The experimental setup is shown in Fig. \ref{fig7}.

\begin{figure}[h!]
\includegraphics[width=0.95\textwidth]{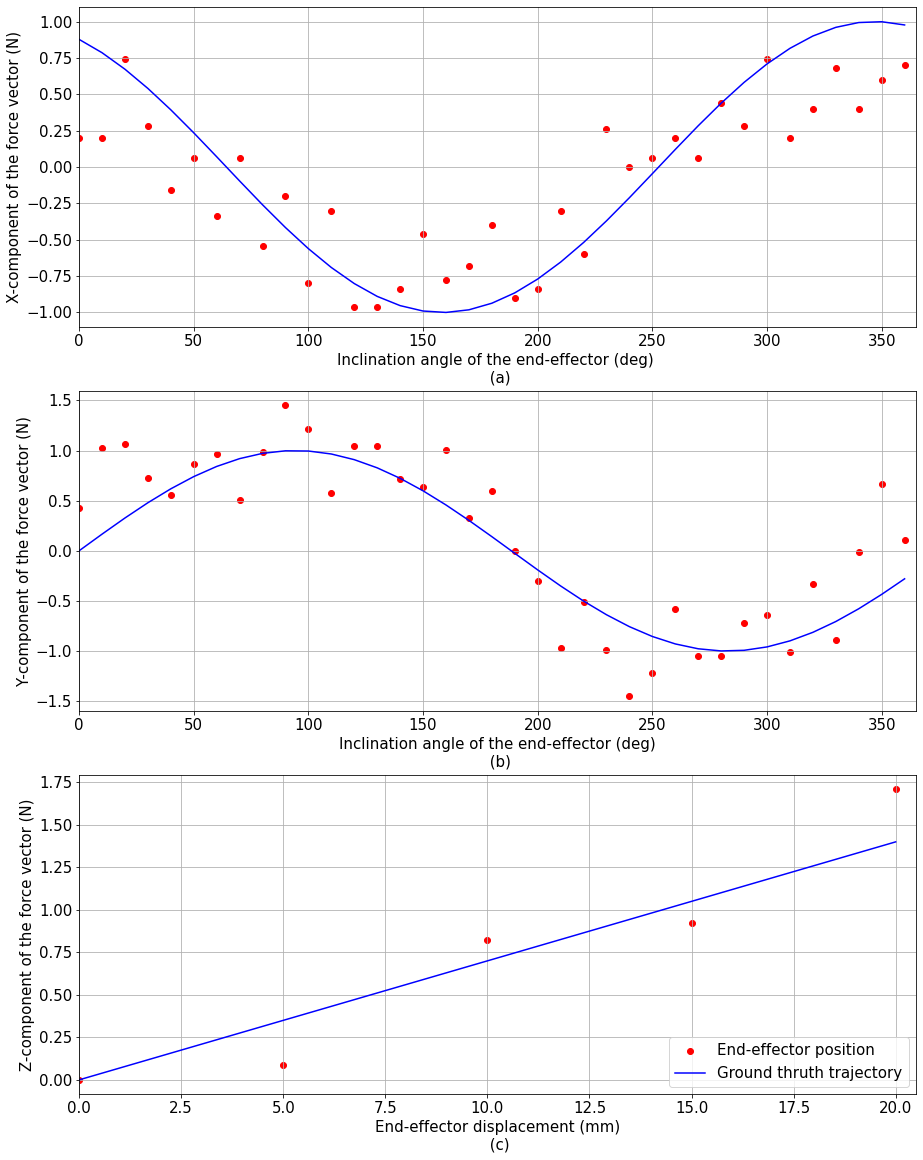}
\caption{Force evaluation experiment.(a) X-, (b) Y-, and (c) Z-components of the force applied by DeltaFinger display alongside the "ground truth" trajectory that is linearly proportional to the end-effector displacement.} 
\label{fig9}
\end{figure}

To evaluate the performance of DeltaFinger, the end-effector was programmed to pass a set of circular trajectories with an increasing radius and distance to the force sensor. The experiment was conducted until the motors were not able to change the end-effector's position further, indicating the highest force feedback. The results of the experiment are shown in Fig. \ref{fig9}. 

% \begin{figure}\centering
% \subfloat[]{\label{a}\includegraphics[width=.7\linewidth]{figures/X Force.png}}\par 
% \subfloat[]{\label{b}\includegraphics[width=.7\linewidth]{figures/Y Force.png}}\par 
% \subfloat[]{\label{c}\includegraphics[width=.7\linewidth]{figures/Z Force.png}}
% \caption{Force evaluation experiment. X (a), Y (b), and Z (c) components of the force applied by DeltaFinger display.}
% \label{fig9}
% \end{figure}

 %For this purpose were measured the X and Y components of the force using a force sensor. 
 With the end-effector rotating by the circle trajectory, the Y and X components of the force vector are defined as the outputs of a sine and a cosine wave functions. The graph shows that the interface performed the trajectory correctly with a maximal amplitude of 0.8 N in transverse plane. In the second experiment, the interface's ability to apply a force along Z axis was tested. The experiment showed that the interface correctly applied forces up to 1.8 N.

% \begin{figure}
% \includegraphics[width=\textwidth]{fig8.png}
% \caption{First experiment. X component of the force} \label{fig8}
% \end{figure}
% \begin{figure}
% \includegraphics[width=\textwidth]{fig9.png}
% \caption{First experiment. Y component of the force} \label{fig9}
% \end{figure}
% \begin{figure}
% \includegraphics[width=\textwidth]{fig10.png}
% \caption{Second experiment. Z component of the force} \label{fig10}
% \end{figure}
\subsubsection{Experimental Results:}
The experimental results suggested that the interface renders the interaction force correctly. The direction and amplitude of the linear force were obtained with the following error (f is the ratio of the mean squares treatment to the mean squares error; p is the probability of error that corresponds to the f-statistic):

\begin{itemize}
 \item The magnitudes of lateral forces ($f_x$ and $f_y$) are kept relatively at the same level with the average standard deviation $\delta = 0.04$ (for normalized force values) and the mean magnitude of 1 N (for original force values).

 \item The deviation of lateral forces did not depend on the diameter of the rotation ($f = 0.11, p = 0.73$ for $f_x$, and $f = 0.43, p = 0.57$ for $f_y$). 
 \item The normal force has a higher standard deviation $\delta$ of 0.16 and mean amplitude of 1.8 N.
\end{itemize}

\subsection{Experiment on Direction Recognition of Linear Force Vector}

We conducted a user study to evaluate the user's perception of a force vector.

\subsubsection{Participants:} We invited 10 volunteers (8 males, 2 females) aged from 21 to 25, right-handed, for the DeltaFinger evaluation. Three participants were familiar with haptic interfaces. Seven participants had never interacted with haptic displays before. 

\subsubsection{Procedure:} All the participants were familiarized with the DeltaFinger force rendering approach prior to the experiment. A blinded experiment was then carried out based on the methodology proposed by Endo et al. \cite{Endo_2010}. The actuator was fixed in at 24 positions on a circle spaced ${\pi}/ {12}$ rad apart, pulling user's index finger in each direction. The performed angles were first demonstrated and then shuffled to be presented in random order to each volunteer. The participant pointed out the perceived direction on the monitor with the patterns displayed on the 360 deg protractor. The main condition of the experiment was that the participant selects the answer based only on haptic feedback.

\subsubsection{Experimental Results:}

\begin{figure}[h!]
\includegraphics[width=\textwidth]{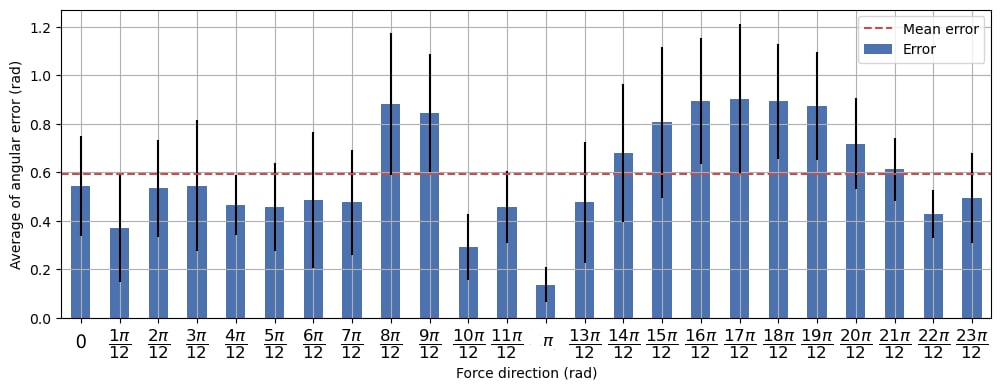}
\caption{Force vector direction recognition error by subjects.} \label{fig10}
\end{figure}

The results of the experiment are shown in Fig. \ref{fig10}.

According to the one-way analysis of variance (ANOVA) with a 5\% significance level, there is a statistically significant difference in user perception of different angles ($F=10.22, p=0.001 < 0.05$). The mean force vector error in user perception is of $0.6$ rad. The results obtained suggest that the user perceives the force transmitted by the haptic interface with a high degree of accuracy. The error in degrees is comparable to the error in force vector perception evaluated in prior haptic research. The difference in error is dependent on the angle due to the peculiarities of tactile perception of the force vector with the index finger. This proves that with the haptic interface, the participants in the experiment were able to obtain high quality information about the direction of the transmitted force. %In reality, when examining an object, a person usually only uses tactile sensations to gain additional information about the object. Also in VR, the developed device will allow additional tactile information about the virtual object.

\subsection{Conclusion}

In this paper, the haptic interface for high-fidelity rendering of force vectors was presented. The distinctive feature of the DeltaFinger interface is the ability to deliver a force interaction experience in VR in nearly omnidirectional space by guiding the fingertip with the delta mechanism. The experimental results suggest that the display accurately renders force vectors with up to 1.8 N amplitude. The results of the user study revealed that users were able to perceive force vectors with a sufficient accuracy (mean error of 0.6 rad).

In future work, we plan to evaluate the ability of the DeltaFinger to render a force vector during contact with virtual surfaces that obtain non-linear stiffness properties. The minimal and maximal transferable stiffness is now also limited by the positional accuracy of the parallel mechanism and servo motor load. Therefore, we are planning to improve the design of the device to allow it to apply a higher range of force amplitudes while preserving the accuracy of the interface. 

The DeltaFinger interface can be potentially applied in virtual scenarios for medical palpation simulators, precise teleoperation, and entertainment, e.g., VR musical applications. Additionally, the device may be potentially applied outside the VR scope for rehabilitation and guidance of people with vision problems.
%This connection to VR opens up additional possibilities for expanding the scope of VR technology. The device allows both the module and the direction of force interaction to be transmitted. Thus, areas of science, technology and communication that require physical force interaction with objects are now available for integration in VR. 

\section{Acknowledgements}
The reported study was funded by RFBR and CNRS, project number 21-58-15006. The authors would like to thank Prof. Sergey Vorotnikov (BMSTU) and PhD Student Daria Trinitatova (Skoltech) for their support of the project.

 \bibliographystyle{splncs04}
 \bibliography{lib}

\end{document}